\documentclass[referee]{aa}
\usepackage{epsfig}
\begin{document} 
\title{A Search for Very Active Stars in the Galaxy 
\thanks{Based on data collected at ESO/VLT Paranal (Chile) Program ID 71.D-0176(A),
Australia Telescope Compact Array (CSIRO), Very Large Array (NRAO, USA),
Anglo-Australian Observatory and Special Astrophysical Observatory (Russia)}
}
\subtitle{First results}
\author {G. Tsarevsky\inst{1,2,3} 
\and Jos\'e A. de Freitas Pacheco\inst{4}
\and N. Kardashev\inst{2}
\and P. de Laverny\inst{4}
\and F.~Th\'evenin\inst{4}
\and O. B. Slee\inst{1}
\and R.~A.~Stathakis\inst{5}
\and E. Barsukova\inst{6}
\and V. Goransky\inst{7}
\and B.~Komberg\inst{2}
}
\offprints{Gregory.Tsarevsky@atnf.csiro.au}
\institute {Australia Telescope National Facility, CSIRO, PO Box 76, Epping, NSW 1710, Australia
\and Astro Space Center, Lebedev Physical Institute, 84/32 Profsoyuznaya St., 117997 Moscow, Russia
\and Isaac Newton Institute of Chile, Moscow Branch, 13 Universitetskij Pr., Moscow 119992, Russia
\and Observatoire de la C\^ote d'Azur, B.P.4229, F-06304 Nice Cedex 4, France
\and Anglo-Australian Observatory, PO Box 296, Epping, NSW 1710, Australia
\and Special Astrophysical Observatory, Nizhnij Arkhyz, Karachaevo-Cherkesia 369167, Russia
\and Sternberg Astronomical Institute, 13 Universitetskij Pr., Moscow 119992, Russia
}
\date{Received date; accepted date} 
\abstract{
We report the first results of a systematic search near the plane
of the Galaxy
for the so called very active stars (VAS), which are characterized by a hard X-ray 
spectrum and activity in the radio domain.
Candidates with hard X-ray binary-like spectra have been selected 
from the Bright ROSAT Source Catalogue in the Zone of Avoidance ($| b | < 20{^o}$) 
and were tentatively identified in GB6/PMM/NVSS radio surveys. Most of them were
observed with the ATCA and VLA. Precise radio coordinates have led to unambiguous optical 
identification for 60 candidates, and a sub-sample of five of them
has been observed with the VLT. 
Also some discovery and confirmatory spectra were obtained with the AAT (4-m) and BTA (6-m). 
Spectroscopy with moderate dispersion, made with the FORS1 spectrograph of the VLT
has revealed two stellar objects (one of them, VASC J1628-41, is definitively
a binary VAS), one new AGN and two featureless spectrum sources. One of these objects, VASC J1353-66, 
shows a marginal evidence of proper motion, which, if confirmed, 
would imply the discovery of a new type of galactic source.
 }
\maketitle

\section{Introduction}
We report the first results of a search for very active stars (VAS) in the Galaxy, 
defined as "ordinary" active stars, i.e., stars detected as 
bright X-ray emitters, having a hard X-ray spectrum and  
activity in the radio domain. These aspects are detailed below and compared 
with other similar active objects.\\
Active stars (AS) are members of a wide class presenting variability
in different spectral domains with prominent emission in X-ray and,
in many cases, in radio. Included in this category are eruptive variables (e.g.,
Orion and UV Ceti type stars); cataclysmic (or explosive) variables (CV) (e.g.,
different types of novae and nova-like stars); RS CVn type and X-ray binaries (XRB) 
(including microquasars, MCQ). 
The latter sub-class, MCQ, will be given special attention in this paper.\\
An important key to understand the nature of AS concerns their distribution in the "two-colour" 
X-ray diagram, HR1 -~HR2. These quantities are hardness ratios resulting from the 
combination of X-ray fluxes in four bands covering the range 
0.1 - 2.0 keV of the ROSAT PSPC instrument (Voges et al. 1999). Such a plot was
exploited by Motch et al. (1998, see their Fig. 2), 
which clearly shows that XRBs (unlike "ordinary" active stars, CV or AGNs) 
have a prominently strong concentration in a narrow strip in 
the extreme upper right corner of the HR1 - HR2 plane. Motch et al. (1998)
claimed that such a concentration could not be produced by selective X-ray absorption in the disk of the Galaxy. 
Therefore, an extreme hardness ratio is a real and intrinsic property of XRB (and MCQ).\\
According to Mirabel \& Rodr\'iguez (1999), microquasars are XRBs displaying highly accelerated radio jets. 
So we can search for new VAS, including MCQ, using X-ray hardness and radio detection as basic criteria 
for selection. In contrast, "ordinary" AS are preferentially soft (and even ultrasoft) X-ray emitters, 
and not all of them are well detected radio emitters. Hereafter, we refer to VAS objects having: 
a) a hard X-ray spectrum (in terms of the ROSAT energy range, 0.1 - 2 keV), 
and b) a detectable radio emission.
To date, approximately 50 objects among the $\sim$ 300 known XRBs (Liu et al. 2001) 
have already been detected
at radio wavelengths (Wendker 1995, Fender \& Hendry 2000, McCormick 2003), but only 
$\sim$ 16 of them have a radio emission morphology typical of MCQs. 
Thus, the present number of known microquasars 
is still too small to perform meaningful statistical studies in order to answer questions concerning their 
nature, evolution and other phenomena.
In spite of the small number of well-classified MCQs, some major (though obviously quite controversial) 
characteristics can be mentioned: a)~a~black hole (BH) or a neutron star (NS) as a primary; 
b)~a~ high mass (HM)
or a low mass (LM) secondary;
c) transient or persistent emission at different wavelengths; and d) in some cases are associated  
with X-ray novae. These characteristics indicate that MCQs constitute a very heterogeneous family,  
stimulating researchers to detect new objects of this class. In particular, 
Paredes et al. (2002) surveyed the plane of the Galaxy in the latitude range 
$|b| < 5{^o}$ in the area covered by the NRAO VLA Sky Survey (NVSS), which is restricted to declinations 
$\delta > -40^o$, i.e., about a quarter of the Milky Way was not covered. 
They did not find any new MCQ and came to the conclusion that the number of MCQ in the Galaxy 
is probably very small (Marti et al. 2004). Our data seem to confirm such a conclusion.

We have extended this search to $|b| < 20{^o}$ along the whole Zone of Avoidance (ZOA), 
aiming to find new very active stars, including MCQs, in the Galaxy.

It is interesting to note that those sources from the Bright ROSAT Source Catalogue (1RXS release) by Voges et al. 
(1999) having $HR1 \geq 0.90$ and $HR2 \geq 0.25$ (i.e., those falling down to the "XRB strip") show a remarkable 
concentration towards the galactic plane and the galactic centre. This sample of 759 objects ( 526 are within
our selection area) consists of different 
types of active sources, and contains an unknown proportion of extragalactic objects. Thus this concentration could be
influenced, at least partially, by selective HI absorption of the X-ray emission. Nevertheless, it 
probably indicates that some fraction of the sources traces a genuine Galactic population of VAS. 

In the present paper we report spectroscopic observations of a sample of five VAS candidates, 
performed with the VLT at ESO (Paranal), supplemented by observations made with 
the Anglo-Australian Telescope (AAT) and the 6-m Russian Telescope (BTA). 
The radio data were obtained with the Australia Telescope Compact Array
(CA) and the Very Large Array (VLA). 
These observations are part of a complete survey conducted in both hemispheres using
selection criteria described in Sec.~2. In Sect.~3 observations and data reduction are described,
and in Sect.~4 we analyze the objects individually. Finally, in Sect.~5 we present our concluding remarks.

\section{The Sample: Selection Criteria}

The idea and corresponding strategy to find new VAS are as
follows (Tsarevsky et al. 2002, 2003)
\footnote{ 
The survey algorithm described below is similar to that one developed 
by Paredes et al. (2002).}:

 Step 1:
Selection of low galactic latitude sources ($|b| < 20{^o}$) from  
the ROSAT All-Sky Survey Bright Source Catalogue, 1RXS release (Voges et al. 1999) 
by the following criteria specific to XRB:

$0.90 \leq HR1 \leq 1.00$, and $0.25 \leq HR2 \leq 1.00$.\\
\noindent It is possible that this selection criterion missed some of the VAS, 
if they were temporarily in a soft state at the time of the ROSAT
observations. But we expect the number of missing objects to be 
relatively small. Indeed, for XRBs, Fig. 2 of Motch et al. (1998) shows that
most of them are in a hard state, which could be explained by comptonization in the 
corona around the accretion disk.
We have also discarded sources with position errors of more than $15\arcsec$ 
and extension parameter more than $100\arcsec$. Thus, we have selected an initial sample
of 428 sources for further study.

 Step 2:
In line with the VAS definition (see above), finding of tentative radio
identifications for the sources of Step 1 in the following surveys: GB6 and 87GB 
(4.8 GHz, Gregory et al. 1996)
\footnote{It is useful to use GB6 and 87GB catalogues together, to check for
the radio variability which is expected to be a strong possibility for sources like VAS.}
, PMN (4.8 GHz, Griffiths et al. 1994) and NVSS (1.4 GHz, Condon et al. 1998). 
An upper limit of the ROSAT to radio coordinate uncertainty was chosen 
as 1.0 arcmin for the 87GB/GB6/PMN, and 0.6 arcmin for NVSS catalogues correspondingly.
This way we tentatively identified in radio 201 of 428 ROSAT sources.

 Step 3:
Further high-sensitivity and sub-arcsec resolution radio observations 
with the CA and VLA, to obtain accurate coordinates for unambiguous optical identifications,
and to find their radio morphology and possible variability.
We have already observed 85 of 201 sources, and five of them are briefly 
reported in this paper.

 Step 4:
Optical identifications on DSS plates, which, due to the precision 
of the CA/VLA coordinates, are in most cases unambiguous       
(i.e., radio-optical coordinate difference, $|r-o|$, is less than $1.5\arcsec$,
see examples in Tab. 2), 
even in overcrowded fields along the Milky Way. For some sources still not 
observed by the CA/VLA, we have also used NVSS data, especially for those brighter 
targets with more precise coordinates (28 in our case). 
Such NVSS-only based identifications are not unambiguous and
need confirmation by high resolution CA/VLA observations.\\ 
At this stage of the survey, we have discarded galaxies clearly recognized in the DSS. 
Note that most of these are new radio galaxies behind the Milky-Way, i.e., in the 
ZOA -- a valuable by-product of this investigation. In this 
way, 53 galaxies have been discarded from further consideration in the context 
of this project. Therefore, only 60 star-like objects were retained for subsequent 
spectroscopic observations.

 Step 5: 
Moderate resolution spectroscopic observations of the 
optically identified star-like VAS candidates (VASC). 
We have already made such observations for 50 VASC
along the whole ZOA using the following telescopes: 
AAT (4-m, AAO, Australia), 
BTA (6-m, SAO, Russia) and VLT UT2 
(8-m, ESO, Chile). \\
After this step, some objects clearly showed their AGN (QSO/Sy) nature
(e.g., object VASC J1757-41 in our current sub-sample, see Sec. 4), 
and therefore they constitute another by-product of the survey. 
Excluding extragalactic objects, the remainder are stars (e.g., 
VASC J1626-33 and VASC J1628-41, see Sec. 4) believed to be 
potential VAS candidates. 

\section{Observations and Data Reduction}

\subsection{Radio Observations}

Following step 3 of our selection algorithm (Sec. 2), we have observed our
VAS candidates using the CA (at 4.8 and 8.6 GHz in a 6-km configuration
during 2000-2003) for the objects south of $DEC = -30^o$, and the VLA
(at 4.8 GHz in A-configuration on 18-Oct-2000) for the only object
in the current sub-sample north of $-30^o$, VASC~J1942+10. \\
The data were routinely processed with the AIPS and MIRIAD packages.
Four of the five sources have not been resolved by the sub-arcsec
resolution of the CA/VLA, so we do not show here their radio maps.
One source, J1757-41, a half-Jy object and a newly discovered Seyfert
(see Tab. 2 and Sec. 4.4), shows arcsec radio structure and a
definitely unresolved core. Thus precise radio coordinates were
measured for these five sources, permitting their unambiguous
optical identifications. Table 2 contains the resulting radio
parameters. \\
More details of the observations and data reduction will be given
in a separate paper, together with radio data of all 85 sources
observed.

\subsection{Optical Identifications}

Thanks to the unambiguous optical identifications mentioned above,
there is no need to show finding charts for the
four of five sources considered in this paper. Indeed,
they can easily be found via a conventional on-line DSS facility
(e.g., USNO B1.0 Catalogue) - see Table 2 for corresponding coordinates
and magnitudes.\\
However, source VASC J1626-33 has a close companion - an anonymous galaxy
barely resolved in the DSS plates. Thus, we have adjusted the slit
orientation to obtain simultaneous spectra of both objects. An excellent
VLT map of this complex, reconstructed from the FORS1 image record, is
shown in Fig. 4 (see Sec. 4.2 for more details).

\subsection{Optical Spectroscopy}

The spectra of five VAS candidates 
were obtained with FORS1 (Focal Reducer low dispersion Spectrograph) 
attached to the 8-m KUEYEN (UT2) telescope, during the period
June-September 2003. The GRISM 600V was used, allowing a wavelength coverage 
ranging from H${\beta}$ to H${\alpha}$ with a spectral resolution of 200
km~s$^{-1}$ FWHM. The slit width was uniformly set to 1$^{\arcsec}$. 
The orientation of the slit for each target was chosen in order to include 
one or two sources close to the center of the radio identification and 
to check possible misidentifications in some
crowded areas. Integration times up to 45 min were scheduled 
to achieve the desired S/N~=~50.
The data were reduced using standard procedures of MIDAS package. 
We made a sky background subtraction
to eliminate undesirable emission lines and a wavelength calibration, 
using He, Hg/Cd and Ar spectra. 
Pixels with a deviation greater than $5\sigma$ from the expected intensity 
profile were ignored. This allowed eliminating the majority of cosmic rays and CCD defects.\\
We have also used spectra obtained with AAT 
in service mode (4-m; RGO spectrograph, 120 to 300 km~s$^{-1}$ resolution, 
typical slit width $1.5\arcsec$) 
and BTA (6-m; UAGS spectrograph, 820 km~s$^{-1}$ resolution, typical slit width $2\arcsec$). 
Details are given in the Table 1.\\
Data were reduced using standard routines, similarly to VLT spectra.

\begin{table*}
\caption{Log of the VLT, AAT and BTA observations and X-ray hardness (HR1, HR2) of the targets.}

\begin{tabular}{llcccccccc}
\hline
VASC       & Name 1RXS        & l   & b  &Epoch          & Telescope & Resolution &$\lambda$-coverage& HR1 & HR2 \\
Name       &                  &deg. &deg.&               &           & (km$s^{-1}$)     &   (\AA)            \\
\hline
$J1353-66$ &J135341.1-664002  &309.1&-4.5& 10-May-03       & AAT       & 120        & $4350-7350$      & 0.93& 0.56 \\
           &                  &     &    & 01/02-Aug-03    & VLT       & 200        & $4855-6960$      &     &      \\
$J1626-33$ &J162620.7-332925  &346.1&10.8& 25/26/31-Jul-03 & VLT       & 200        & $4855-6960$      & 1.00& 0.65 \\
$J1628-41$ &J162848.1-415241  &340.3& 4.7& 02-Apr-01       & AAT       & 170        & $4000-7400$      & 1.00& 0.60 \\
           &                  &     &    & 31-Jul/02-Aug-03 & VLT       & 200        & $4855-6960$      &     &      \\
$J1757-41$ &J175715.6-414914  &350.0&-8.5& 31-Jul-03       & VLT       & 200       & $4855-6960$      & 1.00& 0.61 \\
$J1942+10$ &J194246.3+103339  & 48.3&-6.4& 17-Jun-02       & BTA       & 820        & $5630-8050$      & 1.00& 0.67 \\
           &                  &     &    & 18/23/25-Sep-03 & VLT       & 200        & $4855-6960$      &     &      \\  
\hline
\end{tabular}

\end{table*}

\section{Results}

\begin{table*}
\caption{Results of radio and optical observations}
\label{observations}
\begin{tabular}{lllcrrcccl}
\hline
          &\multicolumn{2}{l}{CA/VLA Coordinates}&NVSS & CA/VLA &\multicolumn{1}{c}{CA} & Sp.  &\multicolumn{2}{c}{USNO} & \\
 VASC     &\multicolumn{2}{c}{(4.8 GHz)}         &F1.4 &\multicolumn{1}{c}{F4.8}&\multicolumn{1}{c}{F8.6}& Ind. &$|r-o|$  & R   & Spectral \\ 
 Name     &RA 2000     & DEC 2000   & mJy &\multicolumn{1}{c}{mJy}&\multicolumn{1}{c}{mJy} &$\alpha$&$\arcsec$& mag & Classification  \\ 
\hline
$J1353-66$&13 53 40.15 &$-66$ 39 57.58& {---}       & 48.2 $\pm 0.1$ &41.7 $\pm 0.1$ &$-0.2$& 0.5     & 17.1& Featureless \\

$J1626-33$&16 26 23.08 &$-33$ 29 33.62& 21.8$\pm 0.8$& 0.7 $\pm 0.1$  &0.8 $\pm 0.1$  &+0.2: & 0.7     & 15.1& Star K0 III \\

$J1628-41$&16 28 47.285&$-41$ 52 39.14& {---}       & 8.8 $\pm 0.1$  &12.3 $\pm 0.1$ &+0.7  & 0.7     & 12.4& Star K3 III-IV\\

$J1757-41$&17 57 15.658&$-41$ 49 18.76& {---}       & 488.5 $\pm 0.2$&398.9 $\pm 0.2$&$-0.3$& 0.3     & 17.0& Sy1, z = 0.3342 \\

$J1942+10$&19 42 47.484&+10 33 27.11  & 99.0$\pm 0.5$& 106.0 $\pm 0.2$&\multicolumn{1}{c}{---}&+0.1: & 0.2     & 16.1& Featureless \\ 

\hline
\end{tabular}

\end{table*}

The results from inspection of our spectra, as well as other related optical and radio data, 
are summarized in Table 2.
Columns 4~- 5 give radio coordinates obtained at 4.8 GHz with the CA
or VLA (for the VASC J1942+10 only). The subsequent four columns 
contain a brief summary of radio data: 
flux densities at 1.4 GHz (taken from NVSS), 4.8 GHz (CA or VLA), 
8.6 GHz (CA) and 
a spectral index $\alpha$
(in the sense $F_{\nu}$ $\propto$ ${\nu}^{\alpha}$), respectively.
Column 8 shows a radio to optical identification quality as a corresponding position 
difference, $|r - o|$.
Column 9 gives R-magnitude taken from the USNO B1.0 Catalogue.
The last column gives a brief spectrum description detailed in the following subsections.\\

\subsection{VASC J1353-66}

VLT data show a featureless spectrum with the interstellar (IS) NaI 
doublet and some weak diffuse interstellar bands (DIB) at $5770\AA$ 
and $6280\AA$ (Fig. 1). The AAT spectrum, obtained on 10-May-2003, 
confirms the featureless spectrum of the object.
Taking into account the detected X-ray and radio
emission, this source is probably a new BL Lac type object behind the Galaxy.
Its position in the sky is in the vicinity of the Great Attractor 
(Woudt \& Kraan-Korteweg 2001).
However, it is worth mentioning that the Naval
Observatory Merged Astrometry Dataset (NOMAD Catalogue) displays a small
proper motion for this object
($\mu(RA) = 20$~$\pm~5$, $\mu(DEC) = -4$~$\pm~6$ mas/y). Despite 
being significant at the 4$\sigma$ level, 
is a marginal result
since only three low-quality Schmidt-DSS plates were measured.
If confirmed, it would suggest a discovery of a new type of
galactic source (see discussion in Sect. 5). 
The resonant doublet Na I$\lambda\lambda$5890-96 is quite well detected
and barely resolved in our VLT spectra, with an intensity ratio between
components
near unity (see Fig. 1) and an estimated equivalent width of EW(D$_1$+D$_2$)~=~2.96\AA. 
The Na I doublet is primarily an indicator of the interstellar reddening
(Munari \& Zwitter 1997) but it can be used to give a rough distance estimate
(Allen 1973). Since the line is saturated, only a lower limit of 3 kpc can
be derived from the statistical relation between distance and EW(NaI). Even so,
this lower limit implies a rather high transverse velocity $\sim$300
km.s$^{-1}$, which cannot presently be excluded for such a peculiar object.

\begin{figure*}[t]
\centering
\includegraphics[width=17.0cm,height=5.0cm]{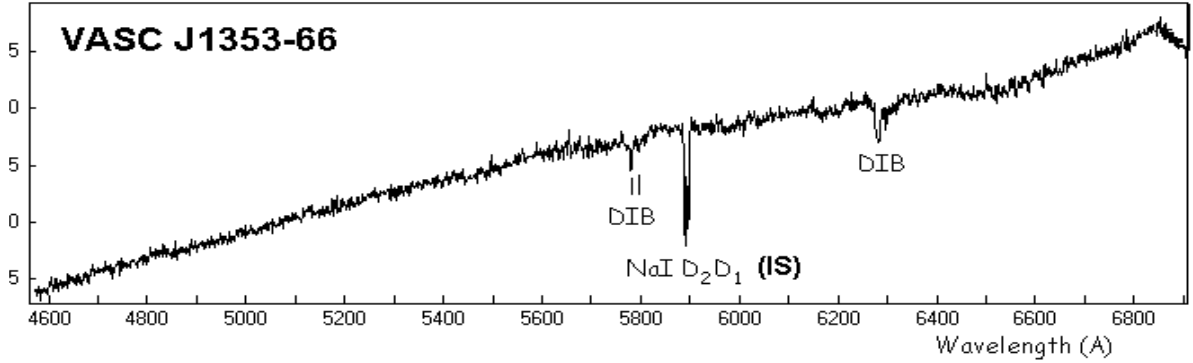}
\caption{Featureless spectrum VLT of VASC J1353-66. Interstellar features indicated:
diffuse bands (DIB) and a NaI doublet.}

\label{figure1}

\end{figure*}


\begin{figure*}[t]
\centering
\includegraphics[width=17.0cm,height=5.0cm]{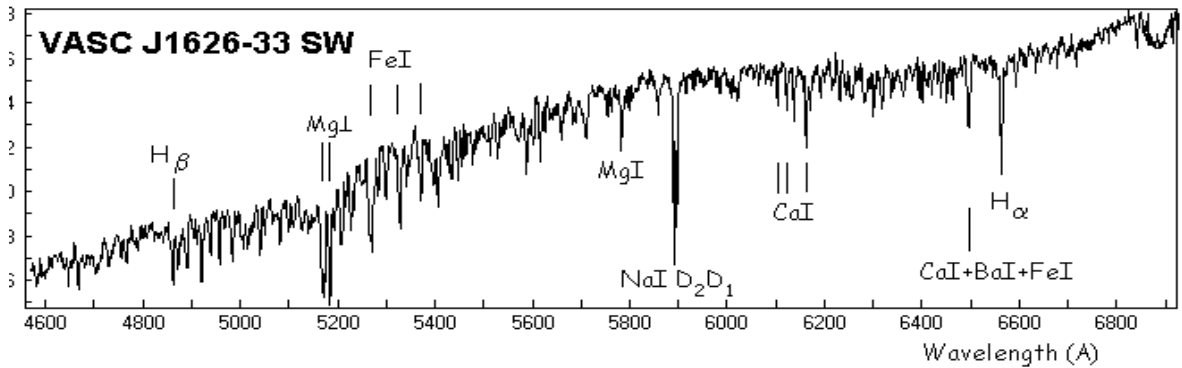}
\caption{VLT spectrum of the SW component of VASC J1626-33, 
identified as a K0 III star.}
\label{figure2}
\end{figure*}

\subsection{VASC J1626-33}

There are two barely resolved objects on the DSS plates: a star at the radio position
(see Table 2) and a galaxy in $3.5\arcsec$ $PA = 45^o$ from the star.
The slit was conveniently orientated in order to include both objects. 

\parindent=3mm SW component: The VLT spectrum indicate that this object is a K0-1 III star, 
classified according to comparison with a template set and using the intensity ratio 
H$\alpha$/(CaI+BaI+FeI), with a radial velocity of $\rm +24 \pm 2 km s^{-1}$.\\  
This object is a potential VAS candidate. Fig. 2 shows the VLT spectrum, 
including identification of main spectral lines.\\
The NVSS catalogue gives at this position a 21.8 mJy source at 1.4 GHz, which contradicts CA
detection below 1 mJy at 4.8 and 8.6 GHz on 03-Jun-2000, suggesting a possible
strong radio variability (see Table 2).

\parindent=3mm NE component: This object was identified as an elliptical galaxy at 
$z = 0.1092 \pm 0.0002$. The classification and
estimated redshift are consistent with the identification of absorption features as the
G-band, H$\beta$, the MgI-blend and the NaI doublet.
Using FORS1 imaging ability, we have constructed a high quality R-image of the field shown in Fig. 4.
Inspection of this image indicates a morphology typical of an early-type
galaxy (E0), confirming 
the interpretation resulting from the analysis of its spectrum
(Fig. 3). Coordinates and magnitude were derived from the image shown in
Fig. 4 as: RA=16h26m23.2s, DEC=-33$^o$29$^m$32.1'' (J2000) and R=17.0m.\\
Both sources possibly have infrared (2MASS) counterparts, 16262293-3329342 and 16262316-3329319,
respectively. High resolution X-ray observations are needed to sort out which
of the two
sources is 1RXS J162620.7-332925, initially selected as a VASC candidate.

\begin{figure*}[t]
\centering
\includegraphics[width=17.0cm,height=5.0cm]{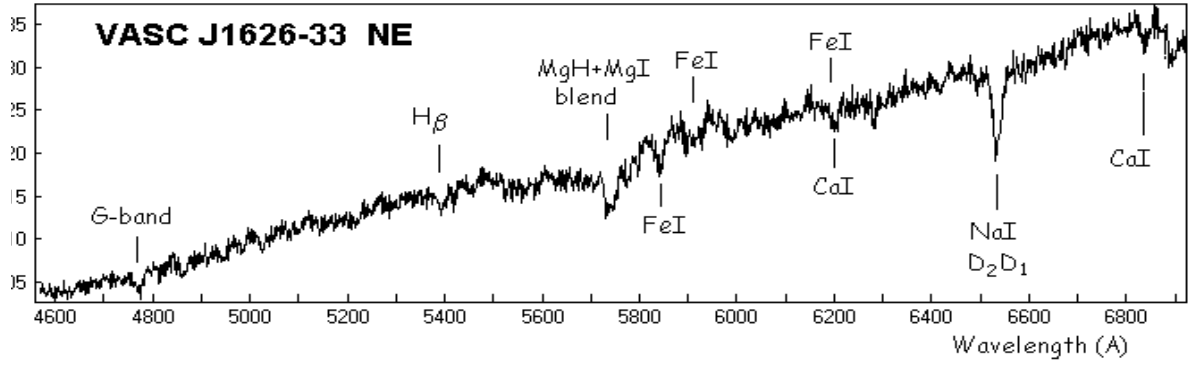}
\caption{VLT spectrum of the NE component of VASC J1626-33, 
identified as an elliptical galaxy E0 at z = 0.1092. See VLT FORS1 optical image in Fig. 4.}
\label{figure3}
\end{figure*}

\begin{figure}[t]
\centering
\includegraphics[width=8.5cm]{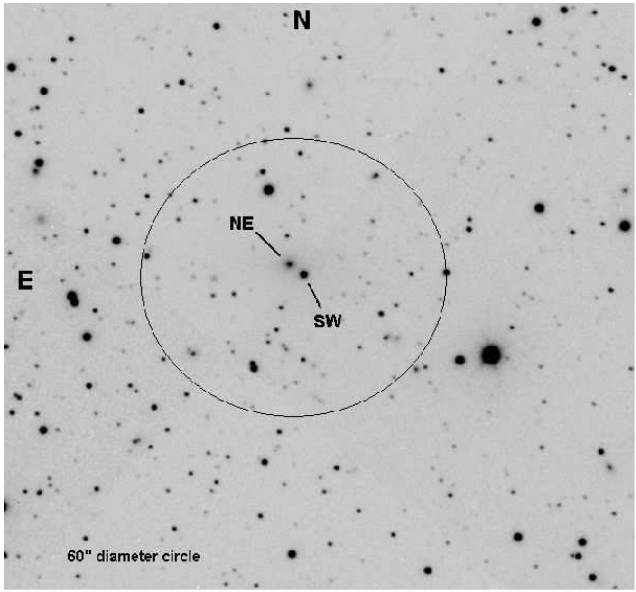}
\caption{FORS1 VLT R image for the VASC J1626-33 complex:
SW (star - VAS candidate) and NE (elliptical galaxy E0 at z= 0.1092); 
$d = 3.5\arcsec$, $PA = 45^o$.}
\label{figure4}
\end{figure}

\subsection{VASC J1628-41}

A relatively bright star, which shows a strong variable H$\alpha$ emission 
discovered in April 2001 by AAT spectroscopy (Tsarevsky et al. 2001). 
In contrast, the VLT spectra show a weak emission and a small variation 
between successive nights (see Fig. 5). Corresponding equivalent widhts, EW, are as follows:

31-Jul-2003 $EW = -1.04 \pm 0.08$ $\AA$ 

02-Aug-2003 $EW = -0.41 \pm 0.10$ $\AA$ \\
\noindent It can be compared with the much stronger $H\alpha$ emission displayed 
in Tsarevsky et al.~(2001) - by retrieving via URL:\\ 
http://www.atnf.csiro.au/people/gtsarevs/$J1628-41$$\_$Halpha.ps 

The spectrum is typical of a late metal-poor giant star. In the Table 1 we adopted a spectral 
classification K3~III-IV made by Torres et al. (2004) using the MIKE echelle
on the 6.5-m 
Baade telescope.\\ 
The binary nature of this star was recently suggested by Buxton et al. (2004) 
and confirmed by Torres et al., who assigned it
to the RS CVn class. However this possibility needs confirmation by additional radial velocity 
data. If this object is in fact a RS CVn, 
it would be the hardest X-ray emitter of this class. 
An inverted radio spectrum (Table 2) and a transient radio emission of VASC J1628-41 
were evidenced through CA and VLA observations (Slee et al. 2002; Rupen et
al. 2002).
Face to the observations, we conclude that this object is definitively a VAS. 

\subsection{VASC J1757-41}

This object is a strong radio source with an unresolved core and a rather
flat spectrum (see Tab. 2), and also a star-like optical object chosen
by our selection criteria. The VLT spectrum (Fig. 6) shows emission lines
typical of AGN (broad Balmer lines and narrow forbidden lines, corresponding to
a redshift of $z = 0.3342 \pm 0.0002$). The full width at zero intensity 
of the $H\beta$ emission is about  12200 km $s^{-1}$, 
consistent with that of a Seyfert 1 galaxy, which in our case is seen through the 
plane of the Galaxy. The forbidden [OIII]$\lambda$5007 lines are quite 
extended, corresponding to about 8.5" along the slit. 
So the projected linear extent of the emitting region at the measured 
redshift is about 44 kpc (for a flat cosmological model with
$\Omega_m$ = 0.3, H$_0$ = 65 km $s^{-1}$ $Mpc^{-1}$).

\begin{figure}[t]
\centering
\includegraphics[width=8.5cm]{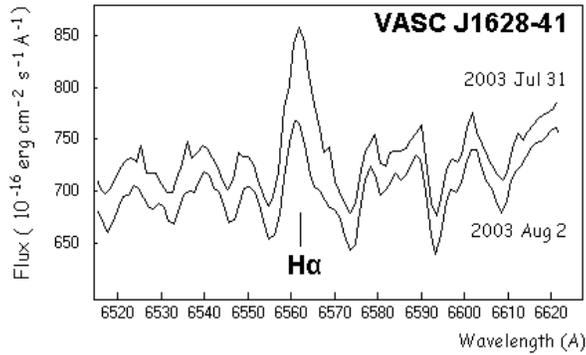}
\caption{VLT spectrum of VASC J1628-41 at the H$\alpha$ vicinity. Notice the H$\alpha$ 
emission variation between two different nights (spectrum of Aug 2 is arbitrary shifted down).}
\label{figure5}
\end{figure}

\subsection{VASC J1942+10}

The VLT data show a featureless spectrum similar to VASC J1353-66, 
with interstellar (IS) NaI 
doublet and some weak diffuse interstellar bands (DIB) at $5800\AA$ 
and $6300\AA$ (Fig. 7). BTA data confirm the featureless spectrum 
of the object. Taking into account the detected X-ray and radio
emission, this source is probably a new BL Lac type object 
behind the plane of the Galaxy.


\begin{figure*}[t]
\centering
\includegraphics[width=17.0cm,height=5.0cm]{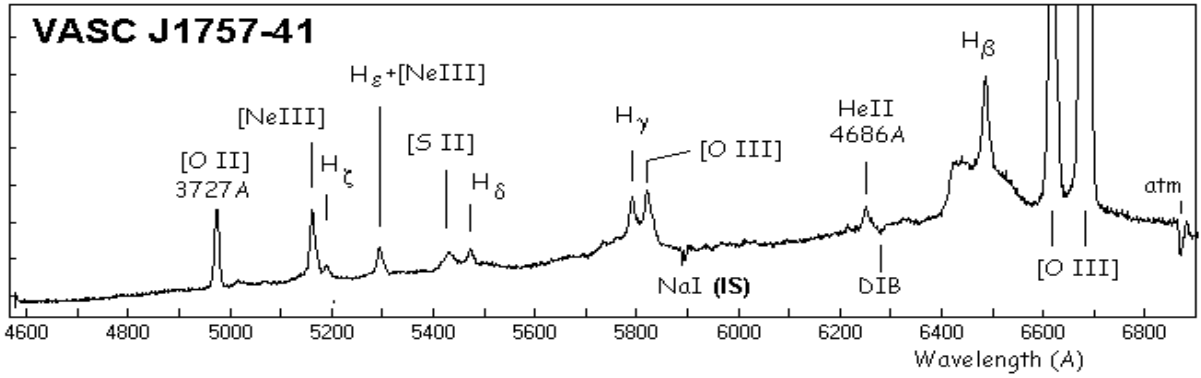}
\caption{VLT spectrum of VASC J1757-41, identified as a Sy 1 at z = 0.3342.} 
\label{figure5}
\end{figure*}

\begin{figure*}[t]
\centering
\includegraphics[width=17.0cm,height=5.0cm]{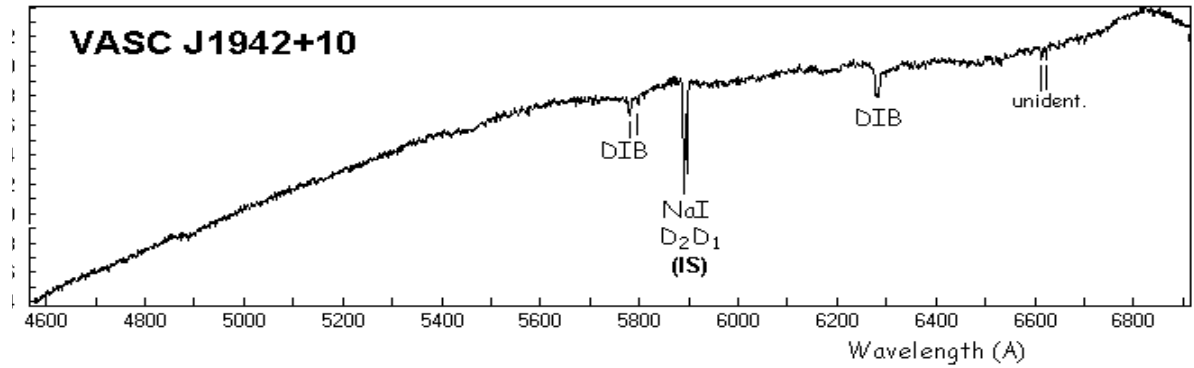}
\caption{Featureless spectrum VLT of VASC J1942+10. Interstellar features are
  indicated:
diffuse bands (DIB) and a NaI doublet. Weak absorption unidentified lines
near $6620\AA$ are also marked.}
\label{figure1}
\end{figure*}

\section{Summary}
In this paper we report the spectral classification of five VAS 
candidates with the VLT-FORS1 spectrograph, using also some of the
AAT and BTA spectra. 
These objects are a part of a complete sample of bright and hard ROSAT sources,
emitting also at radio frequencies,
selected in the plane  of the Galaxy ($\mid b\mid < 20{^o}$). 
Their unambiguous optical identification were possible thanks to
arcsec-resolution observations at 4.8 and 8.6 GHz made with the CA and VLA.

Two of the five candidates are late type giant stars, and therefore  promising 
VAS candidates. Their radio spectral indices (Tab. 2) indicate flat/inverted
radio spectra characteristic of the MCQ family (see Fender 2001). One of two,
VASC J1628-41, due to X ray, radio and optical behaviour, can be definitively
classified as a VAS.
Further optical photometry, high resolution spectroscopy and radio
observations are needed to establish their
variability, binarity, presence of jets, and so on.

Two other objects show featureless spectra, suggesting that they are probably
new BL Lac type
objects in the ZOA (like another two objects discovered in a similar way by Marti et al. 2004).
 Surprisingly, for one of these objects, VASC J1353-66, the NOMAD catalog indicates a small 
proper motion. If confirmed, it would suggest a discovery 
of an interesting new type of Galactic source. Indeed, the only sources in the Galaxy 
having continuous spectra are members of a well known sub-class of white dwarfs (WC), 
which do not display the distinctive radio and X-ray emission required by our
criteria.

Finally, we have also found a Seyfert 1 galaxy (J1757-41) in the ZOA 
at $z =~ 0.3342 \pm 0.0002$. 
The galaxy is also a strong radio source  
and has a star-like appearance in optical images.
This case is very similar to the
well known Seyfert GRS 1734-292 behind the Galactic center, discovered
and investigated by Marti et al. (1998), which was 
initially suspected to be a microquasar.

A statistical study
including all selected and observed VAS candidates will be reported in a separate paper.

\hyphenpenalty1000

\begin{acknowledgements}
We would like to thank E. Budding, A. Burenkov, M. Buxton, L. Berdnikov,
I. Karachentsev, 
W. Orchiston, E. Pavlenko, I. Pronik, M. Rupen, R. Sault and R. Spencer for discussions 
and help with observations. We are indebted L. Staveley-Smith and an 
anonymous referee for his valuable comments, having improved the earlier
version of this paper. 
B.K. and G.T. acknowledge financial support from the Russian Federation
Fundamental Research grant 03-02-16580. G.T. is indebted to colleagues of the 
Cote d'Azur Observatory for their hospitality and stimulating discussions
during his 2003-04 visits. 
We acknowledge use of the SIMBAD, NED, CATS, NVSS, 2MASS and USNO databases 
in the Virtual Observatory regime. The Compact Array is part of the Australia 
Telescope, which
is a National Facility of the Commonwealth of Australia managed by CSIRO.
\end{acknowledgements}
{}

\begin{thebibliography}{}
\bibitem{}
Allen C.W., 1973, Astrophysical Quantities, Athalone Press, p.266 
\bibitem{}
Buxton, M., Bailyn, Ch., Berdnikov, L., \& Tsarevsky, G. 2004, Astronomer's Telegram, ATEL \#234
\bibitem{}
Condon, J. J., Cotton, W. D., Greisen, E. W., et al. 1998, AJ, 115, 1693
\bibitem{}
Fender, R. P. 2001, MNRAS, 322, 31
\bibitem{}
Fender, R. P., \& Hendry, M. A. 2000, MNRAS, 317, 1
\bibitem{}
Gregory, P. C., Scott, W. K., Douglas, K., \& Condon, J. J. 1996, ApJS, 103, 427
\bibitem{}
Griffiths, M. R., Wright, A. E., Burke, B. F. \& Ekers, R. D. 1994, ApJS, 90, 179 
\bibitem{}
Liu, Q. Z., van Paradijs, J., \& van den Heuvel, E. P. J. 2001, A\&A, 368,
1021
\bibitem{}
Marti, J., Mirabel, I. F., Chaty, S., \& Rodriguez, L. F. 1998, A\&A, 330, 72
\bibitem{} 
Marti, J., Paredes, J. M., Bloom, J. S., Casares, J., Rib\'o, M., \& Falco, E. E. 2004, A\&A, 413, 309
\bibitem{}
McCormick, D. G. 2003, PhD Thesis, University of Manchester
\bibitem{}
Mirabel, I.,F., \& Rodr\'iguez, L. F. 1999, ARA\&A, 37, 409
\bibitem[M98]{}
Motch, C., Guillout, P., Haberl, F., et al. 1998, A\&AS, 132, 341
\bibitem{}
Munari, U. \& Zwitter, T. 1997, A\&A, 318, 269
\bibitem{}
Paredes, J. M., Rib\'o, M., \& Marti, J. 2002, A\&A, 394, 193
\bibitem{}
Rupen, M. P., Dhawan, V. \& Mioduszewski, A. J. 2002, IAUC 7968 
\bibitem{}
Slee, O. B., Tsarevsky, G., Sault R., Rupen, M. P., et al. 2002, IAU Circular, 8008
\bibitem{} 
Torres, M. A. P., Garcia, M. R., Steeghs, D., McClintock, J. E., \& Bloom
J. C. 2004, Astronomer's Telegram, ATEL \#292
\bibitem{}
Tsarevsky, G., Kardashev, N., Stathakis, R. A., Slee, O. B., Ojha, R. 2001, Astronomer's Telegram, ATEL \#80
\bibitem{}
Tsarevsky, G., Pavlenko, E., Stathakis, R., Kardashev, N., \& Slee O. B. In: The
Physics of Cataclysmic Variables and Related Objects, ASP Conf. Ser., 
v. 261, eds. B. G\"ansicke et al., 2002, p. 301
\bibitem{}
Tsarevsky, G., Kardashev, N., Slee, O. B., Stathakis, R., \& Budding E. In: New
Directions for Close Binary Studies: THE ROYAL ROAD TO THE STARS,
Publications of Canakkale University Astrophysics Research Center, v. 3, 246, 2003 
\bibitem{}
Voges, W., et al. 1999, A\&A 349, 389
\bibitem{}
Wendker, H. J. 1995, A\&AS, 109, 17
\bibitem{}
Woudt, P. A. \& Kraan-Korteweg, R. C. 2001, A\&A, 380, 441
\end{thebibliography}
\end{document}